\documentstyle[11pt,newpasp,twoside,epsf]{article}
\def\edcomment#1{\iffalse\marginpar{\raggedright\sl#1\/}\else\relax\fi}
\reversemarginpar

\begin{document}
\title{2-D Radiation Transfer Model of Non-Spherically Symmetric Dust Shell in Proto-Planetary Nebulae}
 \author{Kate Y.L. Su, Volk, K. and Kwok, S.}
\affil{Department of Physics \& Astronomy, University of Calgary, Calgary, Canada T2N 1N4}

\begin{abstract}


We have fitted the HST images and the spectral energy distributions (SEDs) of
three proto-planetary nebulae (PPN) with a 2-D radiation transfer model. 
The geometric and mass loss properties of these PPN are also derived.
\end{abstract}

\section{Introduction}

The detection of bipolar reflection nebulosities in PPN suggests that
the dust envelopes in PPN are highly asymmetric.
(see Hrivnak, these proceedings). 
In order to model these objects, a 2-D radiation transfer
model is necessary.  
However, a fully self-consistent determination of the source function in 2-D
is impractical not only because of the large computing time required,
but also in the lack of knowledge of the physical details of the scattering 
process.
Since the morophology of the
nebulae will be primarily determined by the geometry and orientation of 
the envelope, we have developed an approximate solution to 
simultaneously fit the SED and images of a centrally-heated dust envelope.
In this paper, we report the models for 3 PPN, IRAS 17105-3224, 
IRAS 18095+2704 and IRAS 17441-2411, where HST images are available.

\begin{figure}
\plotfiddle{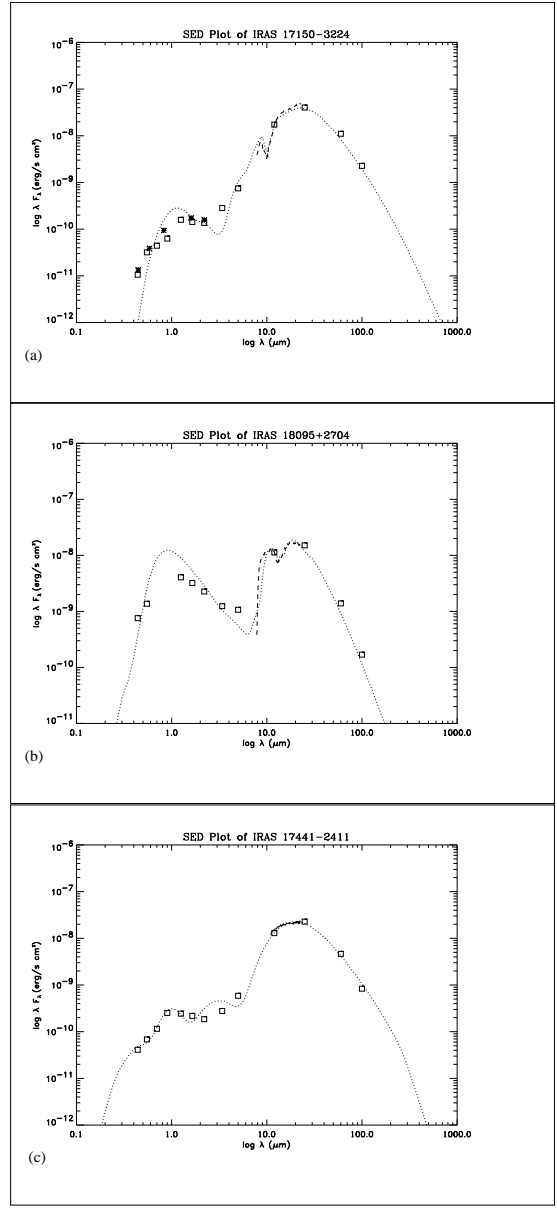}{7 in}{0}{30}{30}{-100}{0}
\caption{The  SED and model spectra for (a) IRAS 17150-3224  
(b) IRAS 18095+2704  (c) IRAS 17441-2411. 
The square and * are ground-based and HST photometric measurements
respectively.  The IRAS LRS is plotted as a dashed line and the model
spectra as dotted lines.}
\end{figure}

\section{The Model}

The dust envelope is assumed to be axial symmetric where the density
distribution $\rho(r,\theta)$ is assumed to have radial cutoffs at 
$r_{in}$ and $r_{out}$. The plane perpendicular to $z$ ($\theta=90^\circ$) 
is referred to as the 
``equatorial plane'' and the directions along the $z$ axis ($\theta=0^\circ$
and $180^\circ$) are referred to 
as the two ``poles''.  The density is assumed to decrease
from the equator to the poles in the form of a power law
($\rho \propto \theta^\beta r^{-\gamma}$).  
In order to produce the searchlight beams observed in
the PPN, IRAS 17150-3224 (Kwok et al. 1998), 
cavities can be put in the density distribution simply assuming a open-cone
structure, where
$\rho$ drops by a factor of $\tau_{scale}$ inside the cone. 
A disk can also be put in the 
density distribution in order to reprduce the dark lane observed in 
IRAS 17441-2411 (Su et al. 1998). 
The viewing angle $i$ is defined as $0^\circ$ if the object is viewed along 
the pole (pole-on), and $90^\circ$ if it is viewed along the equator (edge-on).

We first solve
the dust temperature distribution  at the ``polar" and ``equatorial" directions
from  1-D radiation transfer models.  The values
at other angles are then obtained by interpolation using a power law:

$$ T(\theta) = T(\theta=0) + \left(T(\theta=\frac{\pi}{2})-T(\theta=0)\right)\,
\left[{{2\theta}\over{\pi}}\right]^N\qquad{\rm for~}\theta \leq \pi/2 $$
$$T(\pi-\theta)=T(\theta) $$
The three power indices ($\beta, \gamma, N$) are adjusted until best fits are obtained for both the SED
and the images.  Specifically, $\gamma$ can be
constrained by the infrared color ([60]-[100] $\mu$m) because the total 
amount of dust is fixed.
$\beta$ determines the degree of asymmetry and
is constrained by the observed width of the reflection nebula.
$N$ is basically 
the same as $\beta$, but can be adjusted in order to get the total energy 
conserved.  The viewing angle {\it i} can be determined by comparing the
flux ratio of the two reflection lobes in the simulated image to the one in 
the observed image. 

IRAS 17150-3224 and IRAS 18095+2704 are both O-rich objects, as evidenced  
by the presence of the 9.7$\mu$m silicate feature in the IRAS Low Resolution
Spectra (LRS); therefore, we use silicate dust grains in the fitting.  
The strength of the silicate feature
is also used to constrain the dust opticall depth along
the line of sight.  IRAS 17441-2411 shows no feature in IRAS LRS, we adopted amorphous
carbon dust grains in the fitting.

\section{Results}

Table 1 lists all the fitting parameters we used. 
Figure 1 shows model fits to the 
SED for these three objects. Figure 2 shows the comparision between 
the observed model images. The model images not only reproduce the
approximate shapes of the PPN, but also the absolute flux levels
(as evidenced by the sizes of the outermost contour). 
In addition, 
searchlight beams in IRAS 17150-3224 and dark lane in IRAS 17441-2441
are successfully reproduced as well.

\begin{figure}
\plotone{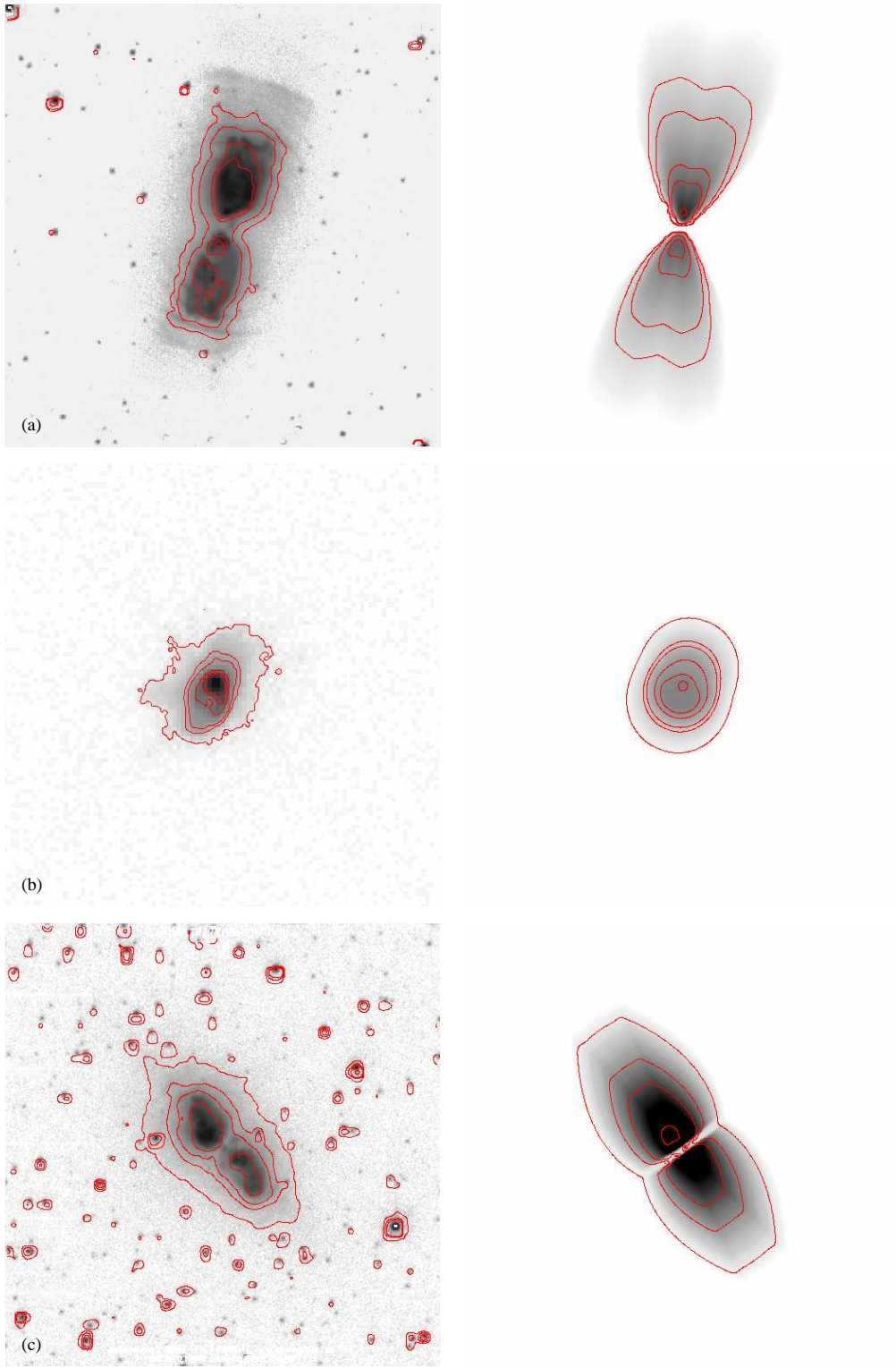}
\caption{Comparision between the observed images and the model ones. 
(a) IRAS 17150-3224  (b) IRAS 18095+2704  (c) IRAS 17441-2411 ~ All the 
grey-scale images are in log. scale, and the contours levels are the same in
both observed and model images for each objects respectively.}
\end{figure}

\begin{table}
\caption{Fitting Parameters}
\begin{tabular}{|c|c|c|c|}
\hline
IRAS& 17150-3224& 18095+2704& 17441-2411 \\ \hline \hline

$i$			& 82	& 60	& 85	\\ 
($\gamma, \beta, N$)	& (2.0, 1.5, 1.0) & (3.5,1.5,1.5) & (2.0,0.5,0.5) \\
$\tau_{pole}, \tau_{equator}$ & 0.3, 4.5 & 0.0736, 0.9 & 0.005, 0.05 \\ 
openning angle of cone& 12$^\circ$	&--	&--	\\ 
$\tau_{scale}$ &	0.005	&--	&--	\\ 
openning angle of disk&--	&--	& $\pm$5$^\circ$ \\ 
density enhancement &--		&--	& 20 \\ \hline
\end{tabular}
\end{table}


\end{document}